\documentclass[runningheads]{llncs}
\usepackage{graphicx}
\usepackage{diagbox}
\begin{document}
\title{An evaluation of U-Net in Renal Structure Segmentation}
%
%
\author{Haoyu Wang\inst{1,2} \and Ziyan Huang\inst{1,2}  \and Jin Ye\inst{2}  \and Can Tu\inst{1,2} \and Yuncheng Yang\inst{1,2} \and Shiyi Du\inst{2,3} \and Zhongying Deng\inst{2,4} \and Chenglong Ma\inst{2,5} \and 
Jingqi Niu\inst{1,2} \and Junjun He\inst{2} \thanks{Corresponding Author}}
\authorrunning{H. Wang et al.}
%
\institute{1 Shanghai Jiao Tong University, Shanghai, China \\
2 Shanghai AI Lab, Shanghai, China \\
3 Sichuan University, Sichuan, China \\
4 University of Surrey, Guildford GU2 7XH, United Kingdom \\
5 Fudan University, Shanghai, China }
\maketitle              
\begin{abstract}
Renal structure segmentation from computed tomography angiography~(CTA) is essential for many computer-assisted renal cancer treatment applications. Kidney PArsing~(KiPA 2022) Challenge aims to build a fine-grained multi-structure dataset and improve the segmentation of multiple renal structures. Recently, U-Net has dominated the medical image segmentation. In the KiPA challenge, we evaluated several U-Net variants and selected the best models for the final submission.

\end{abstract}
\section{Introduction}
Accurate segmentation of renal structures~(e.g. kidneys, renal tumors, arteries, and veins) is the prerequisite of computer-assisted renal cancer treatment. In the preoperative stage, accurate visualization of renal structures can benefit the planning of surgery~\cite{ZHANG201998} and the location of lesions~\cite{SHAO20121001}. In the intraoperative stage, accurate segmentation of renal structures can help guide the clinicians to select arterial clamping branches quickly, thus inducing less collateral damage to the healthy tissues~\cite{NICOLAU2011189}. Therefore, the segmentation of renal structures~(i.e., kidney parsing) is of important clinical significance.

Kidney PArsing~(KiPA 2022) Challenge targets to improve the multi-structure segmentation from abdominal computed tomography angiography~(CTA) for renal cancer treatment. The KiPA challenge is an important step in the development of reliable, valid, and reproducible methods that segment four kidney-related structures on CTA images to promote surgery-based renal cancer treatment\footnote{https://kipa22.grand-challenge.org/home/}. This challenge provides 100+ CTA images with annotations of four target structures: (1) Kidney-Abnormal organ (2) Tumor-Multi-subtype lesion (3) Renal Artery-Very-thin structure (4) Renal Vein-Low-significant region.

Recently, U-Net~\cite{ronneberger2015u} has become \textit{de facto} standard in three-dimensional medical image segmentaion. The design of multi-scale feature extraction and skip connections have been evaluated effective on various segmentaion tasks. Thus, we have evaluated several variants of U-Net in the KiPA challenge and selected the best ones for the final submission.

\section{Method}
\subsection{Network Architecture}
As stated previously, we used two simple variants of U-Net. The vanilla U-Net has an encoder-decoder architecture with multiple skip connections. Specifically, the encoder and decoder can be divided into several stages according to different resolutions of the feature maps. In each stage, both the encoder and decoder stacks two 3D convolution blocks~(i.e. Conv-IN-ReLU) for feature extraction. The first stage of U-Net produces the feature maps with 64 channels, and the numbers of channels is doubled when the stage increases.

All our variants still use 3D convolutions, RELU activation and Instance Norm as the vanilla U-Net. We change the numbers of convolution blocks per stage and the numbers of the feature channels. \textit{Deeper U-Net} has three convolution blocks per stage and 64 initial feature channels. \textit{Thinner U-Net} has two convolution blocks per stage and 32 initial feature channels.

\subsection{Data Augmentation}
We follow the powerful data augmentation settings of nnUNet~\cite{isensee2018nnu}. The key data augmentations are listed in Table \ref{tab:DA_setting}. Different from the default settings in nnUNet, we used bigger range of gamma correction.

\begin{table}[!ht]
    \centering
    \caption{Detailed settings of the data augmentations}
    \label{tab:DA_setting}
    \begin{tabular}{|l|l|}
    \hline
        Data augmentation & Parameters \\ \hline \hline
        Brighness Transform & $\sigma=0.3$  \\ \hline
        Gamma Correction & range=(0.6, 1.6), p=0.3   \\ \hline
        Random Rotation & range=(-60, +60)   \\ \hline
    \end{tabular}
\end{table}

\section{Experiments}
\subsection{Implementation details}
Our development environment is CUDA 11.2, Pytorch 1.11.0 and Tesla A100 80G. We implement our method in the nnUNet framework with the original data preprocessing. The batch size is fixed to 2 and each model is trained for 250, 000 iterations. We use SGD optimizer with momentum of 0.99, weight decay of 0.001. The learning rate is set as 0.01 at the beginning and decayed based on the poly decay policy: $(1-iter/250000)^{0.9}$ for every 250 iterations.

\subsection{Results on the open test set}
The results of different networks trained on the training set~(70 cases) and tested on the open test set~(30 cases) are listed in Table \ref{tab:result}. The deeper U-Net show better performance than other architectures.

\begin{table}[!ht]
    \centering
    \caption{Results of different networks trained on KiPA challenge.}
    \label{tab:result}
    \begin{tabular}{|l|l|l|l|l|l|l|}
    \hline 	 	  	 	  	 	 	  	
        \diagbox{DSC}{Network} & vanilla U-Net & deeper U-Net & thinner U-Net \\ \hline \hline  	 	 	  	 				
        Tumor         &  0.8812 & 0.8926 & 0.8781   \\ \hline
        Kidney        &  0.9545 & 0.9545 & 0.9556    \\ \hline
        Renal Vein    &  0.8354 & 0.8416 & 0.8392    \\ \hline
        Renal Artery  &  0.8726 & 0.8693 & 0.8717    \\ \hline
        Avg.          &  0.8859 & \textbf{0.8895} & 0.8862    \\ \hline
    \end{tabular}
\end{table}

\section{Conclusion}
In this paper, we evaluated three U-Nets on the KiPA challenge dataset~(renal structure segmentation). In our experiments, deeper U-Net shows better performance than vanilla and thinner ones. We hope these empirical results could help further researchers.

%
%
%
\bibliographystyle{splncs04}
\bibliography{ref}
%




\end{document}